\documentclass[aps,twocolumn,superscriptaddress,showpacs]{revtex4} 
\usepackage{graphicx}
\usepackage{amsmath}

\begin{document}

\title{\bf  THE $t$-$J_z$ LADDER}

\author{ A. Wei{\ss}e, R. J. Bursill, C. J. Hamer and Zheng Weihong}
\affiliation{School of Physics, The University of New South Wales,
  Sydney, NSW 2052, Australia}

\date{\today}

\begin{abstract}
The phase diagram of the two-leg t-$J_z$ ladder is explored, using the
density matrix renormalization group method. Results are obtained for
energy gaps, electron density profiles and correlation functions for the
half-filled and quarter-filled cases. The effective Lagrangian velocity
parameter $v_{\rho}$ is shown to vanish at half-filling. The behaviour
of the one-hole gap in the Nagaoka limit is investigated, and found to
disagree with theoretical predictions. A tentative phase diagram is
presented, which is quite similar to the full t-J ladder, but scaled
up by a factor of about two in coupling. Near half-filling a
Luther-Emery phase is found, which may be expected to show
superconducting correlations, while near quarter-filling the system
appears to be in a Tomonaga-Luttinger phase.
\end{abstract}
\pacs{03.70+k,11.15.Ha,12.38.Gc \\[0.5cm] (Submitted to  Phys. Rev. D) }

\maketitle
%\newpage

%\narrowtext
\section{INTRODUCTION}

The discovery of high-temperature superconductivity in the cuprate
materials has sparked huge interest in models of strongly correlated
electrons in low-dimensional systems, such as the Hubbard, $t$-$J$ and
$t$-$J_z$ models. These models are exactly solvable in one dimension, at
least in some special cases; but the two-dimensional models pose a
formidable numerical challenge. The `minus sign' problem is a major
stumbling block for Monte Carlo calculations in these fermionic
systems; and the convergence of Density Matrix Renormalization Group
(DMRG) calculations is slow in two dimensions. Exact finite-lattice
calculations are limited to small lattice sizes; while series
expansions have typically only been useful for special cases such as
the half-filled limit.

In these circumstances, a considerable effort has been invested in the
study of `ladder' systems consisting of two or more coupled chains of
sites. Ladders provide a `half-way house', in some sense, between one
and two dimensions; and they also display some very interesting
effects in their own right \cite{dagotto1996,rice1997}. They display
quite different behaviour depending on whether the number of legs is
even or odd, as in the Haldane effect for the Heisenberg ladders
\cite{haldane1983}. Furthermore, experimental compounds have been
found which form ladders \cite{hiroi1995}, such as SrCu$_2$O$_3$
\cite{azuma1994}, which may allow the theoretical predictions to be
tested experimentally.

The $t$-$J$ model is an `effective Hamiltonian' for the parent Hubbard
model, valid when the Coulomb repulsion is large, but nowadays it is
considered as an interesting model in its own right
\cite{anderson1987,zhang1988}. The $t$-$J_z$ model is a variant in which
the rotational symmetry is broken, and the spin interactions are
Ising-like. The two-leg $t$-$J$ ladder has been extensively studied,
using exact diagonalization
\cite{dagotto1992,tsunetsugu1994,troyer1996,poilblanc1995,hayward1996,hayward1996a,
  mueller1998,riera1999}, quantum Monte Carlo \cite{brunner2001}, the
DMRG technique
\cite{white1997,sierra1998,white1998,rommer2000,siller2001,white2002,schollwock2003},
a combination of different methods \cite{poilblanc2004}, or using
mean-field or approximate analytic methods
\cite{sigrist1994,sierra1998,lee1999,sushkov1999}.  Near half-filling,
the model has been explored using dimer series expansions
\cite{oitmaa1999,zheng2002,jurecka2002}.

Our object is to study the corresponding two-leg $t$-$J_z$ ladder, and
compare the results for the two models.  This model has not been
studied before, as far as we are aware, but the $t$-$J_z$ chain has been
discussed by Batista and Ortiz \cite{batista2000}, and the $t$-$J_z$
model on the square lattice has been treated by several groups
\cite{barnes1989,riera1993,chernyshev1999,riera2001}.  Our primary
tool is the DMRG approach, supplemented with a few series calculations
near half-filling.

The phase diagram for the $t$-$J$ ladder has been discussed by Troyer et
al. \cite{troyer1996}, Hayward and Poilblanc \cite{hayward1996}, and
M{\" u}ller and Rice \cite{mueller1998}. At large $J/t$, the holes all
clump together, and phase separation occurs into hole-rich and
hole-poor regions. At intermediate $J/t$, near half-filling, a `C1S0'
or Luther-Emery phase occurs, where the spin excitations are gapped,
while there is a gapless charge excitation mode.  Troyer et al.
\cite{troyer1996} found evidence of pairing between the holes in this
region, together with long-range superconducting correlations with
modified d-wave symmetry. The spin gap is discontinuous at
half-filling, as the simple magnon excitation gives way to a
particle-hole excitation with spin. At smaller $J/t$, the phase
structure appears to become more complicated, with a possible C2S2
phase appearing (two gapless charge modes and two gapless spin modes)
\cite{mueller1998}; while at extremely small $J/t$, a Nagaoka phase is
expected to appear \cite{nagaoka1966}, where each hole is surrounded
by a region of ferromagnetic spins, forming a ferromagnetic polaron.
In that region no spin gap occurs, and the holes repel each other.

The $t$-$J_z$ ladder might be expected to show similar behaviour. The
major difference between the models is that quantum spin fluctuations
are absent in the $t$-$J_z$ model, and the system exhibits long-range
antiferromagnetic order for half-filling at $T=0$, whereas the $t$-$J$
model does not. This long-range order will be destroyed at any finite
temperature, however, and both models will then display similar
long-range antiferromagnetic correlations. The two models should be
very similar in most other aspects.

This expectation is borne out by our numerical results. The phase diagram
for the t-$J_z$ ladder looks very similar to that of the t-J ladder,
except that the critical couplings are about twice as large, and the
Tomonaga-Luttinger C1S1 phase extends to somewhat higher electron
densities.

In Section \ref{secII} we specify the model, and consider its behaviour
in various limiting cases. In Section \ref{secIII} a brief discussion of
the DMRG method is given, and in Section \ref{secIV} our numerical
results are presented. Our conclusions are given in Section \ref{secV}.

\section{THE MODEL}
\label{secII}

\begin{figure}
  \includegraphics[width=0.7\linewidth]{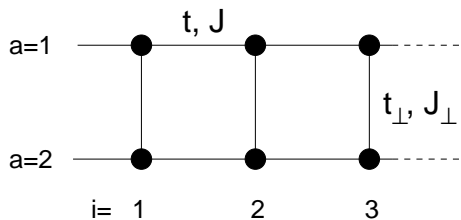}
 \caption{The $t$-$J$ ladder.}
 \label{fig1}
\end{figure}

The Hamiltonian of the $t$-$J_z$ ladder model is (see Figure \ref{fig1})
\begin{eqnarray}\label{tJzham}
H & = & J \sum_{i,a} S^z_{ia}S^z_{i+1,a}
 +J_{\bot}\sum_i S^z_{i1}S^z_{i2} \nonumber \\
 & & -t\sum_{i,a,\sigma} P(c^{\dagger}_{ia\sigma}c_{i+1,a\sigma}
 + h.c.)P \nonumber \\
 & &  -t_{\bot}\sum_{i,\sigma}P(c^{\dagger}_{i1\sigma}c_{i2\sigma} +
h.c.)P
\end{eqnarray}

Here the index $a=1,2$ labels the two legs of the ladder, $i$ labels the
rungs of the ladder, the couplings $J,J_{\bot}$ are the strengths of the
spin interactions on legs and rungs respectively, and $t,t_{\bot}$ are
the hopping strengths on legs and rungs. The projection operators $P$
forbid double occupancy of sites as usual.
A density-density interaction term is sometimes included as a relic of
the parent Hubbard model, but we do not do that here.

\begin{figure}
  \includegraphics[width=0.7\linewidth]{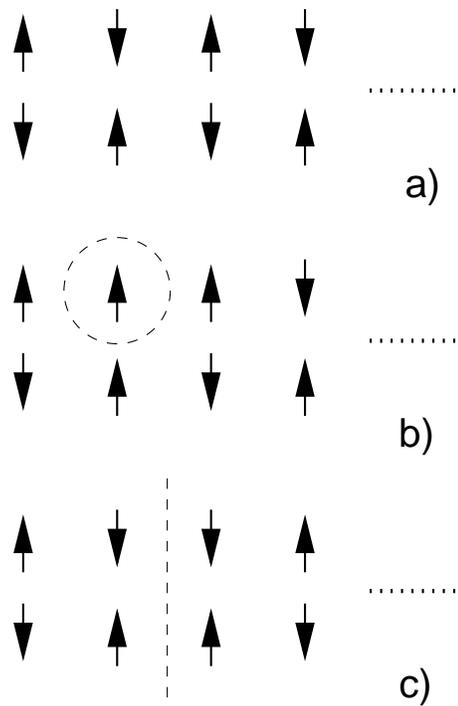}
\caption{Spin configurations at 1/2 filling: a) the antiferromagnetic
ground state; b) an $S^Z = 1$ excitation; c) a domain wall (`soliton') excitation.}
 \label{fig2}
\end{figure}

In the half-filled case, with a single electron occupying every site
(n=1), the model becomes equivalent to a simple classical Ising
antiferromagnet. The ground state is a doubly degenerate
antiferromagnetic state (Fig. \ref{fig2}a), with energy
\begin{equation}
E_0 = -\frac{L}{4}(2J+J_{\bot}),
\label{eq1}
\end{equation}
where $L$ is the number of rungs of the ladder.

The system can be solved exactly in various limiting cases:
\subsubsection{The rung dimer limit ($J/J_{\bot} \rightarrow 0, t/J =t_{\bot}/J_{\bot} $)}

For $J_{\bot} \gg J$, the system consists of independent dimers on the
rungs of the ladder. The eigenstates on a single rung are listed in
Table \ref{table1}. The ground state
is doubly degenerate here, with
both `singlet' and `triplet' states with $S^z = 0$ having the same
energy, unlike the case of the full $t$-$J$ ladder. This degeneracy means
that one cannot simply compute a perturbation series expansion about
the dimer limit in this case, unless one employs degenerate perturbation
theory, or introduces an `artificial' interaction to lift the degeneracy.

\begin{table}
\caption{Rung dimer eigenstates.}
\begin{tabular}{ccccc}
Number & Eigenstate & $S^z$ & Energy & \\ \hline
1 &$ (|\uparrow \downarrow\rangle - |\downarrow \uparrow\rangle)/\sqrt{2}$ & 0 &
-$J_{\bot}/4$ & `singlet' \\
2 & $|\uparrow \uparrow\rangle$  & 1 &
 +$J_{\bot}/4$ &  \\
3 & $(|\uparrow \downarrow\rangle + |\downarrow \uparrow\rangle)/\sqrt{2}$ & 0 &
-$J_{\bot}/4$ & `triplet' \\
4 & $|\downarrow \downarrow\rangle$ & -1 &
 +$J_{\bot}/4$ &  \\
5 & $|00\rangle$ & 0 &
 0 & hole-pair singlet \\
6 & $(|\uparrow 0\rangle + |0 \uparrow\rangle)/\sqrt{2}$ & 1/2 &
-$t_{\bot}$ & electron-hole \\
7 & $(|0 \downarrow\rangle + |\downarrow 0\rangle)/\sqrt{2}$ & -1/2 &
-$t_{\bot}$ & bonding \\
8 & $(|\uparrow 0\rangle - |0 \uparrow\rangle)/\sqrt{2}$ & 1/2 &
+$t_{\bot}$ & electron-hole  \\
9 & $(|0 \downarrow\rangle - |\downarrow 0\rangle)/\sqrt{2}$ & -1/2 &
+$t_{\bot}$ & antibonding \\
\hline
\end{tabular}
\label{table1}
\end{table}

\subsubsection{The independent chain limit ($J_{\bot}/J \rightarrow 0, t/J =
t_{\bot}/J_{\bot}$)}

In this case we end up with two independent chains. The chain
behaviour has been discussed by Batista and Ortiz \cite{batista2000}.
They showed that the spins were N{\' e}el ordered along the chain in
the ground state, which could then be mapped onto an XXZ spin chain.
Phase separation occurs at the (rather large) value $J/t = 8$ for all
hole densities. Below that value, the system forms a gapless Luttinger
liquid, with correlation exponent $K_{\rho}$ and charge velocity
$v_{\rho}$. The system can be exactly solved via the Bethe ansatz, at
quarter filling giving
\begin{equation}
K_{\rho} = \frac{\pi}{4(\pi-\mu)}, \hspace{5mm} v_{\rho} = \frac{\pi t
\sin \mu}{\mu}
\label{eq2}
\end{equation}
where $\cos \mu = -J/8t$. Thus for $0 \leq J/8t \leq
1/\sqrt{2}$, we have $1/\sqrt{2} \leq K_{\rho} \leq 1$, while for
$1/\sqrt{2} \leq J/8t \leq 1$, we have $K_{\rho} > 1$, implying that
superconducting correlations dominate at large distances. At half
filling, on the other hand, we have free fermion (metallic) behaviour
with $K_{\rho}=1/2$.

\subsubsection{The Ising limit ($t/J \rightarrow 0, t/J =
t_{\bot}/J_{\bot}$)}
\label{secII.3}

In this limit, the model becomes equivalent to a classical, static Ising
model. Unless otherwise stated, we assume periodic boundary conditions,
and an even number of rungs L.  The ground state at half-filling is the fully antiferromagnetic
state shown in Figure \ref{fig2}a, with $S^z=0$. The low-energy spin
excitations will consist either of localized flipped spins (`magnons')
with $S^z= \pm 1$, as shown in Fig. \ref{fig2}b, or else domain wall
(`soliton')
excitations, as shown in Fig. \ref{fig2}c, analogous to the `spinon'
excitations in Heisenberg chains, but carrying $S^z=0$ in this case.
The total number of solitons, in the absence of holes, and assuming
periodic boundary conditions, has to be
even for L even, and odd for L odd, as in the Heisenberg chain. The
energy gaps for these two excitations are respectively
\begin{eqnarray}
E_{2b)} - E_0 & = & J + \frac{J_{\bot}}{2} \nonumber \\
E_{2c)} - E_0 & = & J
\label{eq3}
\end{eqnarray}

Since there are no quantum spin fluctuation terms in the Hamiltonian,
these spin excitations are of course static, in the absence of holes.
Note that since the solitons carry integer spin, there is no
possibility of spin-charge separation in this model.

\begin{figure}
  \includegraphics[width=0.7\linewidth]{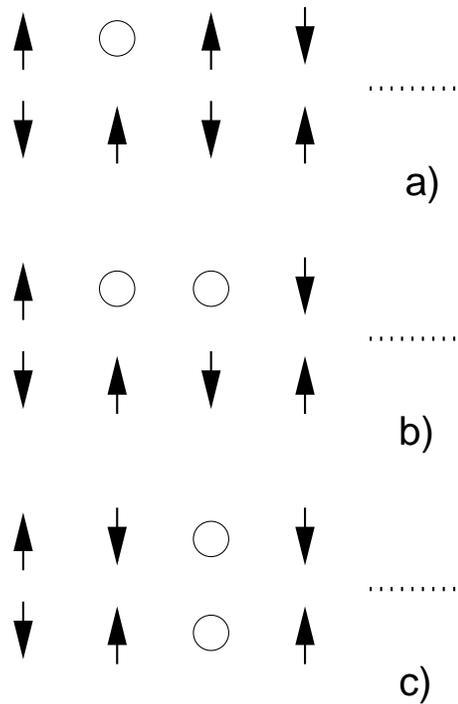}
\caption{Charge excitations near to 1/2 filling: a) a 1-hole state;
b),c) 2-hole states.}
 \label{fig3}
\end{figure}

The lowest-energy charge excitation will consist of a single hole in
the antiferromagnetic background (Fig. \ref{fig3}a), with spin $S^z=
\pm 1/2$, and energy
\begin{equation}
E_{3a)} - E_0  = \frac{1}{4}(2 J +J_{\bot})
\label{eq4}
\end{equation}
The lowest eigenstates in the 2-hole sector consist of a bound pair
on adjacent sites (Figs. \ref{fig3}b,\ref{fig3}c), with spin $S^Z = 0$
and energies
\begin{eqnarray}
E_{3b)} - E_0 & = & \frac{1}{4}(3J + 2 J_{\bot}) \nonumber \\
E_{3c)} - E_0 & = & \frac{1}{4}(4J + J_{\bot})
\label{eq5}
\end{eqnarray}

It is clear that holes will cluster together to minimize the number of
`broken bonds' and hence the energy, and
the system will be phase separated in this Ising limit.

\begin{figure}
  \includegraphics[width=0.7\linewidth]{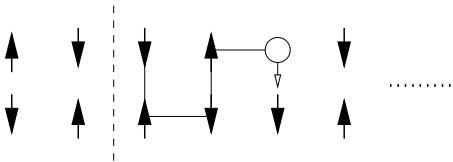}
\caption{A hole-plus-domain-wall excitation, showing an allowed
zig-zag path for hopping (solid line).}
 \label{fig4}
\end{figure}

For small but finite $t/J$, one can study the system via perturbation
series calculations in $t/J$. Some results of these calculations
will be shown in later sections. The single hole states illustrated in Fig.
\ref{fig3}a are localized states, because there is an
energy barrier preventing them from hopping: any hop will disturb the
antiferromagnetic alignment of the spins. At higher energy, however,
there will be states such as that shown in Figure \ref{fig4}, with
unperturbed energy
\begin{equation}
E_{4)} - E_0  =  \frac{1}{4}(8J +J_{\bot}),
\label{eq6}
\end{equation}
where the hole is free to hop along the zig-zag path shown without
incurring any further penalty in spin interaction energy. Assuming
periodic boundary conditions, these states are made up of a hole plus a
soliton for L even, or a hole alone for L odd. For finite $t/J$,
they will form a band of itinerant electron states. The mobility of the
electrons in both the chain and the ladder systems is a major point of
distinction from the two-dimensional model, where the holes are
`confined', i.e. cannot move without creating a `string' of overturned
spins behind them, and paying a penalty in spin interaction
energy \cite{brinkman1970}.
The states (3a) and (4) will mix as soon as $t$ is turned on, and both
configurations should be regarded as hole-soliton bound states.

\begin{figure}
\includegraphics[width=0.8\linewidth]{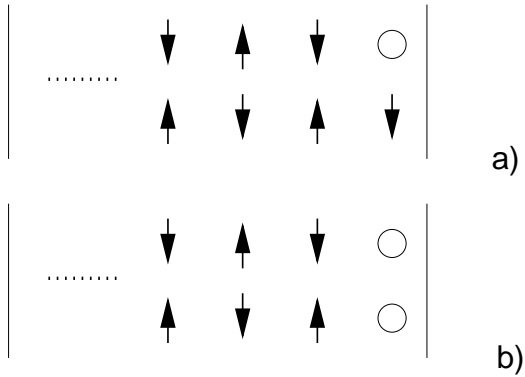}
\caption{Charge excitations for the ladder with open boundaries: a) 1
hole; b) 2 holes.
}
 \label{fig4a}
\end{figure}

We will also employ open boundary conditions in our calculations. In
this case, the holes will cluster towards the boundaries in the Ising
limit, in order to minimize the number of broken bonds. In this case the
ground-state energy is

\begin{equation}
E_0 = -\frac{1}{4}(2J(L-1) - LJ_{\bot}),
\label{eq6a}
\end{equation}
and the lowest 1-hole state (Fig. \ref{fig4a}a) has energy
\begin{equation}
E_{5a)}-E_0 = \frac{1}{4}(J+J_{\bot}),
\label{eq6b}
\end{equation}
while the lowest 2-hole state (Fig. \ref{fig4a}b) has energy
\begin{equation}
E_{5b)} - E_0 = \frac{1}{4}(2J + J_{\bot}).
\label{eq6c}
\end{equation}
The single hole state is not necessarily accompanied by a soliton in
this case.

\subsubsection{The free-fermion limit ($J/t \rightarrow 0, t/J =
t_{\bot}/J_{\bot}$)}

In this limit, a single hole will move through the lattice as a free
particle, and its dispersion relation is naively expected to be
\begin{equation}
E_1(k) = -2t\cos k -t_{\bot}
\label{eq7}
\end{equation}
For simplicity, we shall restrict our remarks henceforth to the
isotropic case $t_{\bot}=t$, when the 1-hole energy gap in this limit
is $-3t$.

The behaviour at small $J/t$ should be similar to that of the $t$-$J$
model. The 1-hole energy in the two-dimensional $t$-$J$ model has been
discussed by a number of authors
\cite{brinkman1970,shraiman1988,kane1989,barnes1989,white2001}.  At
extremely small $J/t$ (of order $10^{-2}$, in the region of the
Nagaoka phase \cite{nagaoka1966,white2001}), the hole will be
surrounded by a ferromagnetic `polaron', or region of
ferromagnetically aligned spins, with spin energy of order $JR^2$ for
a disc of radius $R$, within which the hole is fully mobile. In this
regime, the 1-hole energy is
\begin{equation}
E_{1h}/t \sim -4 + c_1(\frac{J}{t})^{1/2}
\label{eq8}
\end{equation}
where $c_1$ is a numerical constant. At larger $J/t$, however, the
lowest-energy states correspond to Brinkman-Rice `string' states
\cite{brinkman1970,shraiman1988}, where the hole is confined by a
`string' of overturned spins within an antiferromagnetic background.
This picture gives a 1-hole energy
\begin{equation}
E_{1h}/t \sim -a_1 + b_1(\frac{J}{t})^{2/3}
\label{eq9}
\end{equation}
where $a_1 < 4$. Shraiman and Siggia \cite{shraiman1988} estimated $a_1 =
2\sqrt{3}$ and $b_1=2.74$ for the $t$-$J_z$ model, quite close to the
values found numerically by Barnes et al \cite{barnes1989}, and White and
Affleck \cite{white2001}.

For a one-dimensional chain, no Nagaoka phase occurs, but for a ladder
the condition is once again met that holes can hop around closed loops,
and a Nagaoka phase is expected \cite{troyer1996}.
White and Affleck \cite{white2001} have pointed out that the
ferromagnetically aligned spins inside the polaron will orient
themselves in the xy plane, rather than along the $z$ axis, because this
costs an exchange energy of $J/4$ per bond, rather than $J/2$.
For our ladder model, the ferromagnetic `polaron' region will be
one-dimensional. If the polaron region spans L rungs of the ladder, the
cost in magnetic energy will be
\begin{equation}
E_M = \frac{3}{4}JL
\label{eq10}
\end{equation}
Assuming the 1-hole wavefunction to vanish at the edges of the polaron,
it will take the form
\begin{equation}
|\psi_k\rangle = A\sum_n \cos(kn)(|n,1\rangle+|n,2\rangle), \hspace{5mm} k=\frac{\pi}{L}
\label{eq10a}
\end{equation}
where $|n,a\rangle$ denotes a hole at rung $n$ on leg $a$ of the ladder, and
$A$ is a normalization constant. The kinetic or hopping energy is then
easily found to be
\begin{equation}
E_K = -3t + t(\frac{\pi}{L})^2 \hspace{5mm} (L \rightarrow \infty)
\label{eq10b}
\end{equation}
Minimizing the total energy $E_{1h} = E_M + E_K$ with respect to $L$
we find
\begin{eqnarray}
  E_{1h}/t & = & -3 + \frac{3}{4}(\frac{3\pi J}{t})^{2/3} \nonumber \\
  & = & -3 + 3.35 (J/t)^{2/3}
  \label{eq10c}
\end{eqnarray}
For intermediate $J/t$, the Brinkman-Rice string
picture \cite{brinkman1970,shraiman1988} will give a
 qualitatively similar behaviour
\begin{equation}
E_{1h}/t \sim -a_2 + b_2(\frac{J}{t})^{2/3}
\label{eq11}
\end{equation}
where $a_2 < 3$.
It is doubtful whether approximations such as those used by Shraiman
and Siggia \cite{shraiman1988} to calculate the coefficients $a_2$ and
$b_2$ for the 2D model are at all accurate for the ladder, and we will
not attempt an explicit calculation of these coefficients.

\subsection{The Effective Hamiltonian}

In the regime of physical interest, the $t$-$J$ ladder is believed to
be in a C1S0 or Luther-Emery phase, with gapped spin excitations and
gapless charge excitations corresponding to bound hole pairs. Several
authors \cite{troyer1996,hayward1996,siller2001,white2002} have
discussed an effective Hamiltonian to describe these bosonic
excitations, which would capture the low-energy physics of the model.
Here we merely summarize their results.

In the recent analysis of White, Affleck and Scalapino \cite{white2002},
they use a bosonization technique to construct the low-energy effective
Hamiltonian

\begin{equation}
H -\mu L = \frac{v_{\rho}}{2} \int dx [K_{\rho}\Pi_{\rho}^2 +
\frac{1}{K_{\rho}}(\frac{d\theta_{\rho}}{dx})^2]
\label{eq11a}
\end{equation}
where $\mu$ is a chemical potential, $\Pi_{\rho}$ is the momentum
density conjugate to $\theta_{\rho}$, $v_{\rho}$ is the velocity of the
corresponding gapless low-energy excitations and the parameter
$K_{\rho}$ controls the correlation exponents. The two parameters
$v_{\rho}$ and $K_{\rho}$ must be extracted from numerical data.

White et al. \cite{white2002} show that the finite-size scaling behaviour
of a general low-energy excitation is
\begin{equation}
E-E_0 = -2p\mu +\frac{2\pi v_{\rho}}{L}[K_{\rho}m^2 +
\frac{p^2}{4K_{\rho}} +\sum_{k=1}^{\infty}k(n_{Lk}+n_{Rk})]
\label{eq11b}
\end{equation}
where $E_0$ is the ground-state energy for a given density $n$, $n_{Lk}$
and $n_{Rk}$ are occupation numbers for left and right moving states of
momentum $2\pi k/L$, $L$ is the number of rungs of the ladder, and $p$
and $m$ are integer-valued quantum numbers. The total charge relative to
the ground state is $Q=-2p$, and the other quantum number $m$ measures
the ``chiral charge".

By measuring the ground-state energies for three different charge states
with $\Delta Q= \pm 2$, one can determine the ratio $v_{\rho}/K_{\rho}$:
\begin{equation}
E(p=1) +E(p=-1) -2E_0 = \frac{\pi v_{\rho}}{K_{\rho}L}
\label{eq11c}
\end{equation}
This is directly related to the electron compressibility $\kappa$ of the
two-leg ladder, generally defined as
\begin{equation}
\frac{1}{n^2\kappa} \equiv \frac{1}{2L}\frac{d^2E}{dn^2} =
\frac{\pi v_{\rho}}{2K_{\rho}}
\label{eq11d}
\end{equation}
using (\ref{eq11c}), where $n$ is the electron density. This formula
applies for either periodic or open boundary conditions, in principle.
The velocity may be measured independently using the excitation energy of
the lowest state of momentum $2\pi/L$ for periodic BC
\begin{equation}
E(n_{R1}=1)-E_0=\frac{2\pi v_{\rho}}{L}
\label{eq11e}
\end{equation}
For open BC, the momentum spacing is $\pi/L$, and the corresponding
formula is
\begin{equation}
E(n_{1}=1)-E_0=\frac{\pi v_{\rho}}{L}
\label{eq11f}
\end{equation}
Hence one can obtain estimates for $K_{\rho}$ and $v_{\rho}$ separately.

One may also use `twisted' boundary conditions with the wave function
acquiring a phase $\Phi$ at the boundary, corresponding to an
Aharonov-Bohm flux threading the one-dimensional ring formed by
joining the ends of the ladder together. This increases the ground-state
energy by
\begin{equation}
E_0 \rightarrow E_0+\frac{8\pi v_{\rho}K_{\rho}\Phi^2}{L}
\label{eq11g}
\end{equation}
and allows one to determine the combination $v_{\rho}K_{\rho}$. This
approach was used by Hayward and Poilblanc \cite{hayward1996}.

Finally, White et al. \cite{white2002} discuss how one may estimate
$K_{\rho}$ from the decay of Friedel oscillations in the system.
Friedel oscillations are density oscillations produced near the boundary
of an open ladder, which decay with a power law corresponding to the
exponent $K_{\rho}$ with distance into the ladder. Using their
bosonization analysis and a conformal transformation, White et
al \cite{white2002} show that for a system of length $L$ the density at
site (rung) $j$ should vary as
\begin{equation}
\langle n_j\rangle \rightarrow \frac{C\cos(2\pi nj + \beta)}{[(2L/\pi)\sin(\pi
j/L)]^{K_{\rho}}}
\label{eq11h}
\end{equation}
where $C,\beta$ are constants, and $2\pi n$ is the minimum wave vector
of Friedel oscillations in the C1S0 phase, corresponding to two holes
per wavelength. This is the same wave vector that would occur for a
one-component spinless hard core bose gas, made up of tightly bound hole
pairs.

Schulz \cite{schulz1999} has argued in the case of the $t$-$J$ ladder
that $K_{\rho}$ should take the universal value $K_{\rho}=1$ at
half-filling; but his argument starting from the rung dimer limit
is not necessarily applicable to the $t$-$J_z$ model.

Note that the picture may be complicated by the occurrence of a
gapped charge density wave (CDW) phase at special
commensurate filling factors such as $n=0.5$ or $n=0.75$. In this case,
the charge density oscillations persist at all distances from the walls,
corresponding to broken translational symmetry and a regular pattern
of hole placements. This possibility has been discussed by Riera et
al \cite{riera2001} and White et al \cite{white2002}.

\section{METHOD}
\label{secIII}

The model (\ref{tJzham}) is solved using the Density Matrix
Renormalisation Group (DMRG) method \cite{white_dmrg}. Calculations
have been performed at both half-filling and quarter-filling.
Calculations have been performed with both periodic and open
boundary conditions in the horizontal direction, but most of the
results use OBCs because the convergence is much better in this
case.

\begin{figure}
  \includegraphics[width=\linewidth]{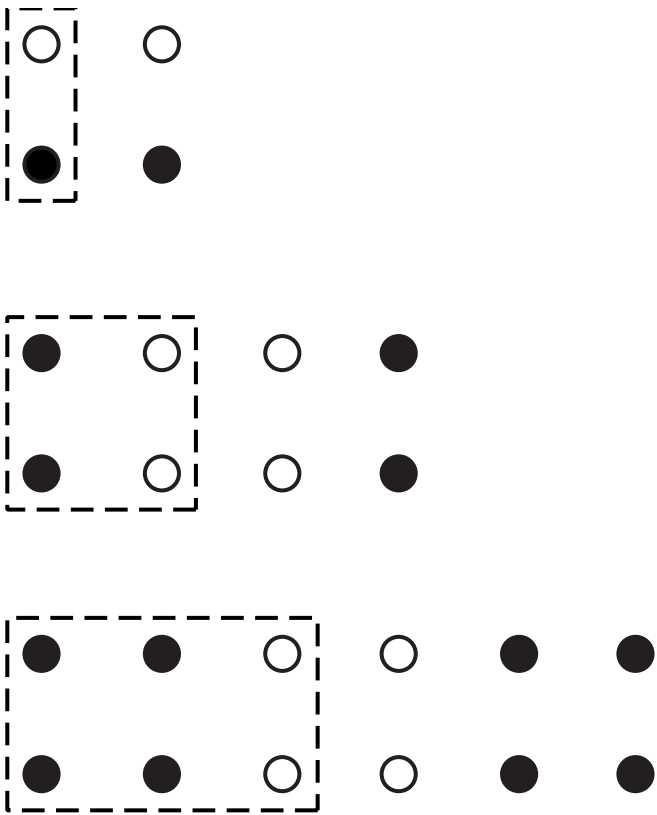}
\caption{Infinite lattice construction of the two-leg ladder. Two
whole rungs (four sites---open circles) are added per DMRG step.
The system and environment blocks increase by one rung (two sites)
per DMRG step.}
\label{infinite_lattice}
\end{figure}

In the half-filled case we have calculated the ground state as
well as one, two and four-hole excited states. In the one-hole
case we have calculated $S^z = 1/2$, 3/2 and 5/2 states. We have
calculated energies of these states as well as density profiles
$\left\langle \hat{n}_i \right\rangle$.
The infinite lattice algorithm \cite{white_dmrg} is employed in
this case, using typically $M = 60$ states per block and symmetry
sector, which corresponds to a total of about 300 system block states.
The lattice build-up phase is illustrated in Fig.\
\ref{infinite_lattice}. By running tests for a number of different
values of $M$, we established that the energies of these states
have been resolved with sufficiently high accuracy for our
purposes.

In the quarter-filled case, it is necessary to use the finite
lattice method \cite{white_dmrg} in order to achieve reasonably
converged energies. We use the infinite lattice method to build a
ladder to a given size, increasing the system block size by a
whole rung at each DMRG step as in Fig.\ \ref{infinite_lattice}.
We then perform a finite lattice sweep at the fixed lattice size,
using the previously developed system blocks to improve the
accuracy of the calculation. Once a target size has been completed
with a finite lattice sweep, the infinite lattice algorithm is
resumed to proceed to a larger target lattice size whereupon the
sweeping procedure is again employed. This way we obtain accurate
finite lattice method results for a number of lattice sizes, e.g.,
$N = 4$, 8, 12, 16, 24, 32, 64, 128, 256 lattice sites.
Wherever necessary, the data are extrapolated to the bulk limit $N
\rightarrow \infty$ by fitting a low-order polynomial in $1/N$ to the
finite-lattice data.

\section{RESULTS}
\label{secIV}

We have calculated numerical DMRG results for this model for the
isotropic case, $J = J_{\bot} \equiv J_z, t = t_{\bot}$, and for various 
$J_z/t$, on
lattices of L rungs with L even. Our discussion henceforth will be limited 
to this
case. 

\subsection{Near Half-Filling}

We begin by considering states with a finite number of holes doped into
the half-filled system (n=1).

\subsubsection{Single-hole states}

\begin{figure}
  \includegraphics[width=\linewidth]{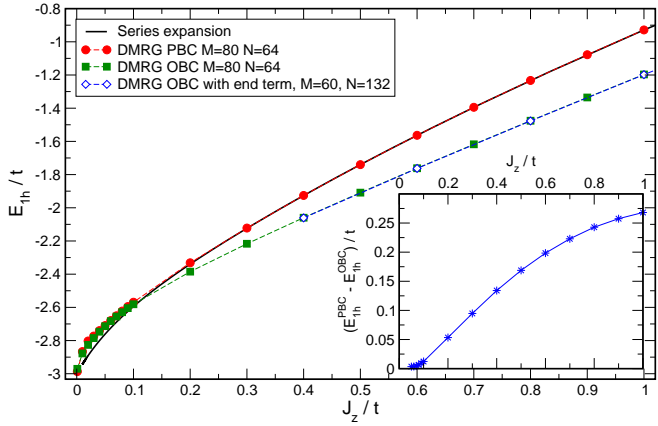}
  \caption{(Colour online) The energy gap $E_{1h}/t =(E_1-E_0)/t$ for
    one hole, as a function of $J_z/t$. Filled circles, open boundary
    conditions; open squares, periodic boundary conditions; dashed
    line, estimates from Ising series expansion.}
 \label{fig6}
\end{figure}

Figure \ref{fig6} shows DMRG estimates of the energy gap for a single
hole with spin $S^z = 1/2$, as a function of $J_z/t$, for both
periodic and open boundary conditions. 
The results of a series calculation are also
shown for the periodic case. The series estimates were obtained assuming
the Nagaoka form (\ref{eq10c})
\begin{equation}
f(t) \equiv E_{1h}+3t \sim bt^{1/3} \hspace{5mm} as \hspace{5mm} t
\rightarrow 0.
\end{equation}
Accordingly, the series in $t$ for $f(t)$ was Euler transformed
$z=t/(1+t)$ to bring the singularity to $z=1$; then a further change of
variable was made
\begin{equation}
1-\delta = (1-z)^{1/3}
\end{equation}
so that one may treat the function $(1-\delta)f(\delta)$ as analytic in 
$\delta$ near $\delta=1$; and finally, differential approximants were
used to extrapolate this function to $\delta=1$. It can be seen that the
series results agree very well with the DMRG results for $J-z/t > 0.3$,
but run a little lower below that.

It can also be seen that there is a difference between the DMRG
results with periodic and open boundary conditions. As discussed in
Section \ref{secII.3}, this is because the state with periodic
boundary conditions is actually a hole-soliton bound state.  The
difference between the two energies is shown in the inset to Figure
\ref{fig6}. 

\begin{figure}
  \includegraphics[width=\linewidth]{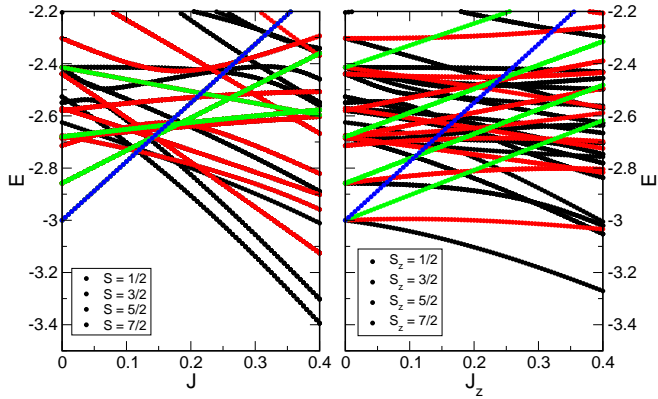}
  \caption{(Colour online) Low energy spectrum of the $t$-$J$ (left)
    and $t$-$J_z$ (right) model on a 4-rung ladder with 7 electrons.}
%   \caption{ The energy gap $E_{1h}/t =(E_1-E_0)/t$ for one hole, as a
%     function of $J_z/t$.}
  \label{fig8}
\end{figure}

At smaller $J_z/t$, one is tempted to look for a transition to a
Nagaoka phase \cite{nagaoka1966}, i.e. a phase with ferromagnetic
spin order or at least a ferromagnetic polaron bubble around the hole.
In the $t$-$J_z$ model, however, the situation is rather different 
from the $t$-$J$ model, where the full spin-rotation symmetry applies.
The
scenario is illustrated in Figure~\ref{fig8}, where we show the
complete 1-hole energy spectrum of the $t$-$J$ and $t$-$J_z$ models on a
4-rung cluster. 
The left-hand panel shows the $t$-$J$ case, where at $J=0$ the ground
state has maximal spin $S=7/2$ (ferromagnetic). As $J$ increases, at
some critical coupling $J_c$ this state crosses another state with
minimal spin $S=1/2$, which becomes the ground state thereafter. The
right-hand panel shows the $t$-$J_z$ case. 
At $J_z=0$, the system is rotation symmetric, the Nagaoka theorem applies
and the ground state again has the
maximal possible spin $S_{\text{max}}$ and degeneracy
$2S_{\text{max}}+1$. As soon as $J_z$ becomes finite, however, the symmetry is
broken, the multiplet splits into its $S^z$ components, and the state
with minimal $S^z = \pm 1/2$ becomes the ground state.
Clearly, there is no level crossover
in the $t$-$J_z$ case. All we can expect for
increasing $J_z$ is a continuous fading-out of the ferromagnetic
correlations.
This accords with the smooth approach of the 1-hole energy to $-3$ as
$J_z \rightarrow 0$ in
Figure \ref{fig6}.

A comparison of the numerical data at small $J_z/t$ with the theoretical
prediction (\ref{eq10c}) gives the following results.
A Dlog Pade analysis of the series  gives a rather
inaccurate estimate of the exponent in the range 0.5 - 0.7, which would
be consistent
with 2/3. A fit to the DMRG data over the range $0-0.1$, however, gives 
\begin{equation}
\frac{E_{1h}}{t} = -3 + 1.444\left( \frac{J_z}{t} \right)^{0.53},
\end{equation}
with an exponent much closer to $1/2$ than $2/3$. This is illustrated by
the difference between the DMRG data and the series extrapolation in
Figure \ref{fig6} at small $J_z/t$. Thus the data do not seem to behave
in accordance with theoretical expectations in this instance.

\begin{figure}
  \includegraphics[width=\linewidth]{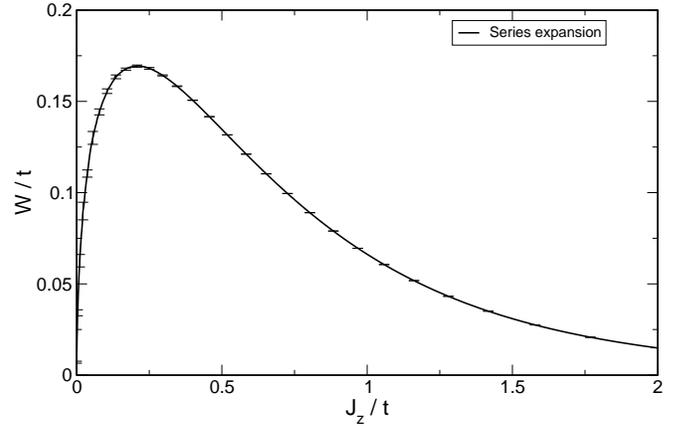}
  \caption{ The 1-hole bandwidth $W/t$ as a function of $J_z/t$, estimated
    from the Ising series expansion.
  }
 \label{fig10}
\end{figure}

Figure \ref{fig10} shows the bandwidth for the 1-hole state with $S^z
= 1/2$ predicted by the Ising series expansion, as a function of
$J_z/t$.  We see that the predicted bandwidth rises as $J_z/t$
decreases, as one might expect, until at $J_z/t \simeq 0.3$ it reaches
a peak, and begins to drop towards zero.  Similar behaviour in the
region of small $J/t$ has been predicted for the 2D $t$-$J$ model by
Martinez and Horsch \cite{martinez1991} and Liu and Manousakis
\cite{liu1992}, and confirmed by series calculations \cite{hamer1998}.
The theoretical predictions assume long-range antiferromagnetic order,
and construct an effective Hamiltonian in which the spin dynamics are
treated in linear spin-wave theory, and the holes are treated as
spinless fermions, following Schmitt-Rink, Varma and Ruckenstein
\cite{schmitt-rink1988}. The results are consistent with the string
picture of Brinkman and Rice \cite{brinkman1970}, and appear to show a
sequence of narrow `string' excitation peaks in the spectral function,
corresponding to holes bound in a linear potential. The effects are
even more marked in the $t$-$J_z$ model.

It would be very interesting to check the actual behaviour
of the bandwidth by means of DMRG estimates, but
unfortunately our codes cannot distinguish different momentum
eigenstates.

\subsubsection{Two-hole states}

In studying 2- and 4-hole states and looking for the phase separation boundary,
we wanted to avoid the holes sticking to the boundary and thus obscuring
the bulk physics. Accordingly, we have added an extra `boundary
potential' term at the sites adjacent to the boundary:

\begin{equation}
H_B = -\frac{J_z}{4} \sum_{i,\sigma} c_{i,\sigma}^{\dagger}c_{i,\sigma}.
\label{eq12b}
\end{equation}

As it turns out, this makes a negligible difference to the energy gaps
in the region of interest.

\begin{figure}
  \includegraphics[width=\linewidth]{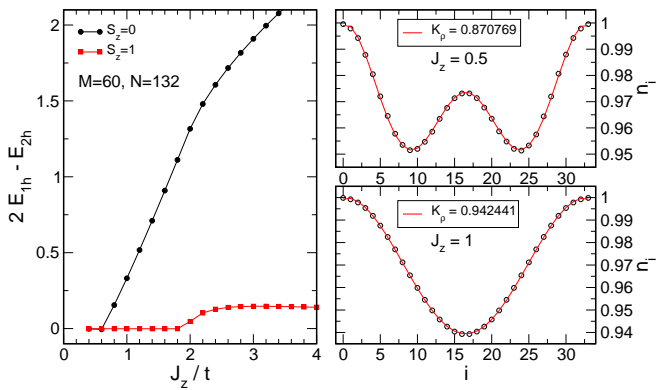}
  \caption{(Colour online) Left: Two-hole binding energies as a
    function of $J_z/t$, calculated with OBC and finite boundary
    potentials.  Filled circles: $S^z = 0$: filled squares: $S^z = 1$.
    Right: Average electron number per rung as a function of distance along the
    ladder, for 2-hole states on a lattice of 34 rungs: a) $J_z/t =
    0.5$; b) $J_z/t = 1.0$. The lines are fits using equation
    (\ref{eq12a}). }
  \label{fig11}
\end{figure}

In the 2-hole sector, the first quantity of interest is the 2-hole
binding energy
\begin{equation}
E_B = 2E_1-E_2-E_0
\label{eq12}
\end{equation}
Figure \ref{fig11} shows the binding energy as a function of $J_z/t$
for both $S^z = 0$ and $S^z = 1$, computed with open boundary
conditions. It can be seen that a 2-hole bound state with $S^z = 0$
forms for large $J_z/t$, but the binding energy vanishes at $J_z/t
\simeq 0.60$, and below this point the holes do not bind. In the $S^z
= 1$ channel, there is a small binding energy at large $J_z/t$, which
vanishes at $J_z/t \simeq 1.8$.

Similar results have been found for other members of this family of
models. For the 2D $t$-$J_z$ model, Riera and Dagotto \cite{riera1993}
estimated that the holes become unbound at $J_z/t \simeq 0.18$; while
for the $t$-$J$ ladder, Jurecka and Brenig \cite{jurecka2002} suggest
that hole binding vanishes at $J/t \simeq 0.5$.  There is binding in
the $S = 1$ channel for the $t$-$J$ ladder, but it is considerably
smaller than in the $S = 0$ channel.

Wherever the 2-hole binding energy is positive, we expect a continuous
band of 2-hole bound states to appear above the ground state; so that
within the 2-hole sector, particle-hole excitations (`excitons') can
occur down to zero energy. In other words, excitons with $S^z = 0$ are
gapless, as found by Troyer et al \cite{troyer1996} for the $t$-$J$
ladder.

We note that the spin energy gap between the $S^z = 1$ and $S^z = 0$
2-hole states also drops to zero at $J_z/t = 0.6$. It always remains
below the $S^z = 1$ `magnon' energy $3J/2$; or in other words, an
exciton with $S^z =1$ in the 2-hole sector has an energy substantially
below that of the magnon. Therefore the spin gap drops discontinuously
as one moves away from half-filling, which again agrees with the
behaviour found by Troyer et al. \cite{troyer1996} for the $t$-$J$ ladder.

The right panels of Figure \ref{fig11} show profiles of the electron
density in the 2-hole state as a function of distance along the
ladder, calculated for a lattice of 34 rungs with open boundary
conditions. At low $J_z/t$, the two holes are unbound and separated
from each other, forming two distinct troughs in the density profile
(Fig. \ref{fig11}). At $J_z/t \simeq 0.6$ a sharp crossover takes
place, where the two holes bind together and form a single trough in the
density profile. As it happens, both profiles can be fitted quite well
with a Friedel oscillation form
\begin{equation}
\langle n_j\rangle = c_0 + \frac{c_1\sin[2\pi (n + 1/2)j/(L +1)]}
{[(L+1)\sin[\pi j /(L+1)]]^{K_{\rho}}}
\label{eq12a}
\end{equation}
where $n=2$ and $K_{\rho}=0.87$ in case a) and $n=1$ and $K_{\rho}=0.92$
in case b).

\subsubsection{Four-hole states}

\begin{figure}
  \includegraphics[width=\linewidth]{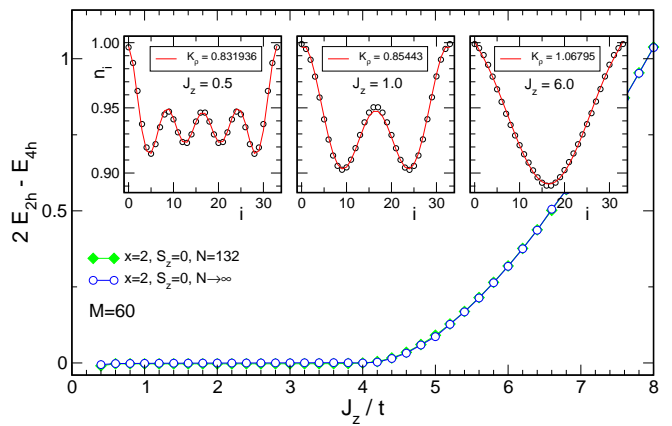}
  \caption{(Colour online) Four-hole binding energy as a function of
    $J_z/t$, calculated with OBC and finite boundary potentials.
    Insets: Electron density profiles at characteristic values of $J_z/t$,
    together with fits to the corresponding Friedel expression.}
  \label{fig13}
\end{figure}

The 4-hole binding energy for $S^z = 0$ is shown as a function of
$J_z/t$ in Figure \ref{fig13}. It is finite at large $J_z/t$, and
vanishes smoothly at $J_z/t \simeq 4$. This is taken to mark the point
where phase separation occurs. The smooth behaviour of the binding
energy may indicate a Kosterlitz-Thouless transition at this point.
The insets show examples of the electron density profiles as functions of
distance along the ladder for the 4-hole ground state. For $J_z/t =
0.5$, we see four separated holes; for $J_z/t = 1.0$, we see two
separated hole-pairs; and for $J_z/t = 6.0$, we see a single 4-hole
cluster, reinforcing the picture given above. 
Note that the Friedel form (\ref{eq12a}) again fits the density profiles
remarkably well, upon choosing the appropriate value for $n$.
The fits in the pair binding region indicate that the parameter
$K_{\rho}$ lies approximately in the range 0.8 - 1.0 near
half-filling.

\begin{figure}
  \includegraphics[width=\linewidth]{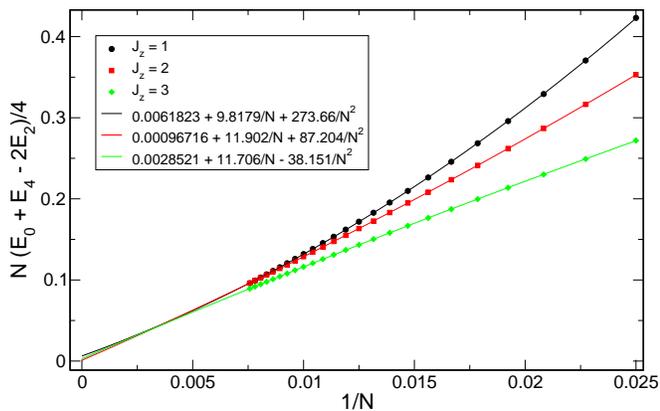}
  \caption{(Colour online) Four-hole binding energy as a function of $1/N=1/(2L)$, at $J_z/t
    = 1$, $2$ and $3$.}
  \label{fig14}
\end{figure}

Next, we attempt to measure the isothermal compressibility in the
non-phase-separated region, whose inverse is
\begin{equation}
  \kappa^{-1} = n^2 N [E_4+E_0 -2E_2]/4
\label{eq13}
\end{equation}
where $N=2L$ is the number of sites, and $n=(N-2)/N$ the electron
density.  Figure \ref{fig14} shows $\kappa^{-1} n^{-2}$ as a function
of $1/N$, for some couplings $J_z/t <4$, fitted by a quadratic
polynomial in $1/N$. The extrapolations to the bulk limit $N
\rightarrow \infty$ are compatible with zero, indicating that the
inverse compressibility vanishes at half-filling. This seems
surprising at first sight, but the same result applies for the 1D $t$-$J$
chain at the integrable point \cite{ogata1991}.
 Siller et al.
\cite{siller2001} also found a tiny result for the inverse
compressibility of the isotropic $t$-$J$ ladder near half-filling.
The result can be understood as follows. The dispersion relation in the
conduction band flattens out near the band edge, and hence the
dispersion relation of a single hole at half filling is quadratic in
momentum $q$, as illustrated by the free fermion equation
\ref{eq7} or the calculation of Sushkov \cite{sushkov1999} for the
t-J ladder. The dispersion relation for a bound hole-pair bosonic
excitation will likewise be quadratic in $q$. The vanishing of any term
linear in $q$ implies that $v_{\rho} \rightarrow 0$ and
Eq.~\eqref{eq11d} then leads to $\kappa^{-1} n^{-2} = 0$.

\subsection{Quarter Filling}

Next we consider states at or near the quarter-filling point ($n=1/2$).
Reliable results are very hard to obtain for $J_z/t \geq 3$. Phase
separation presumably occurs in this region, so that the ground state
becomes inhomogeneous, and very difficult to treat by DMRG methods. For
the most part, our results will be for smaller $J_z/t$.

\subsubsection{Odd-even gap}

\begin{figure}
  \includegraphics[width=\linewidth]{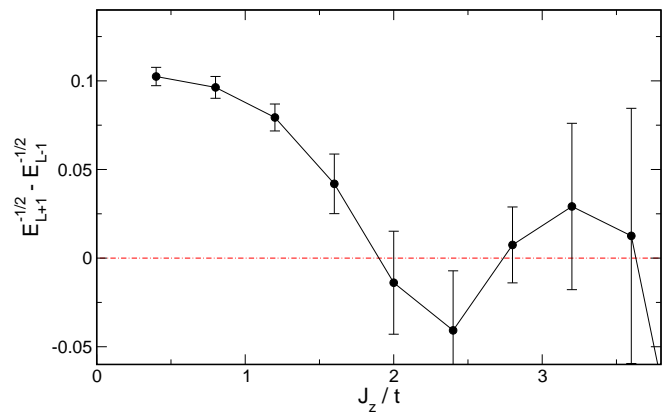}
  \caption{ The odd-even gap
    $[E(n_h=L+1,S^z=-1/2)+E(n_h=L-1,S^z=-1/2)-2E(n_h=L,S^z=0)]/2$ at
    quarter-filling, as a function of $J_z/t$. }
  \label{fig15}
\end{figure}
Figure \ref{fig15} shows the energy gap between states with odd and even
numbers of holes, $[E(n_h=L+1,S^z=-1/2) +
E(n_h=L-1,S^z=-1/2)-2E(n_h=L,S^z=0)]/2$, where $n_h$ is the number of
holes, as a function of $J_z/t$. It can be seen that for $J_z/t$ less
than about $2.0$, 
this gap is small and positive, but is consistent with zero beyond that
point, within substantial errors. This behaviour is very similar to the two-hole gap shown in
Figure \ref{fig18}. In fact it appears that the energy surface is quite
smooth in this region, with little if any `corrugation' between states
with even or odd numbers of holes. This indicates that pairing effects
between holes are weak or nonexistent in this region.

\subsubsection{Spin Gap}

\begin{figure}
  \includegraphics[width=\linewidth]{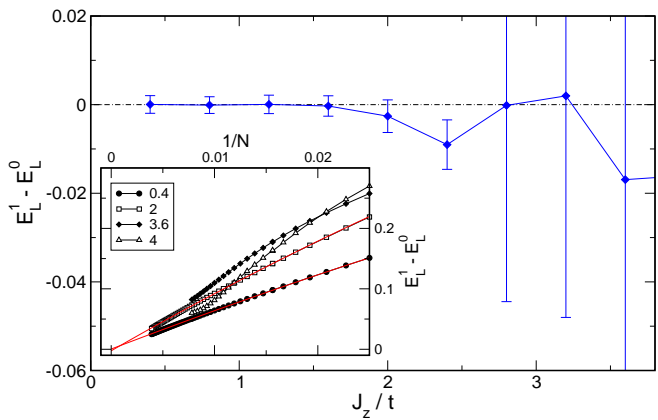}
  \caption{ The spin gap $[E(n_h=L,S^z=1)-E(n_h=L,S^z=0)]$ at
    quarter-filling, as a function of $J_z/t$.} 
  \label{fig16}
\end{figure}

Figure \ref{fig16} shows the spin gap $[E(n_h=L,S^z=1)-E(n_h=L,S^z=0)]$
as a function of $J_z/t$. 
The insert shows examples of the finite-size scaling behaviour of this
quantity as a function of $1/N$, and the linear extrapolations made to
the bulk limit.
It can be seen that the results are consistent
with zero over the whole region $J_z/t
\leq 4$. Convergence for $J_z/t > 2$ is not very good.
Thus for $J_z/t \geq 2.3$,
we appear to be in the C1S1 phase discussed by Hayward and
Poilblanc \cite{hayward1996} for the t-J ladder, where both the charge
gap and the spin gap vanish.

\subsubsection{Two-hole gap}

\begin{figure}
  \includegraphics[width=\linewidth]{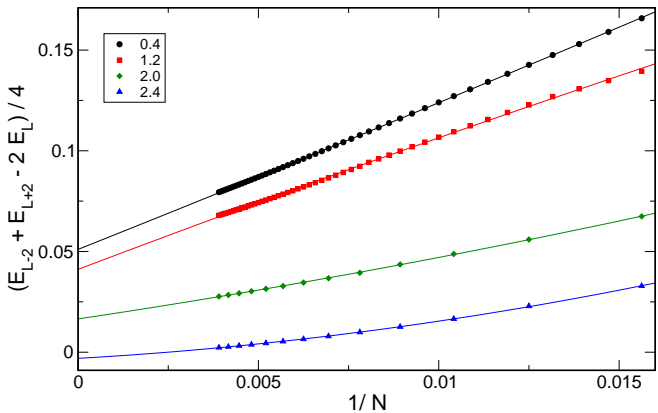}
  \caption{The two-hole gap
    $[E(n_h=L+2,S^z=0)+E(n_h=L-2,S^z=0)-2E(n_h=L,S^z=0)]/2$ at
    quarter-filling, as a function of $1/N = 1/(2L)$, at selected values of $J_z/t$.} 
  \label{fig17}
\end{figure}

\begin{figure}
  \includegraphics[width=\linewidth]{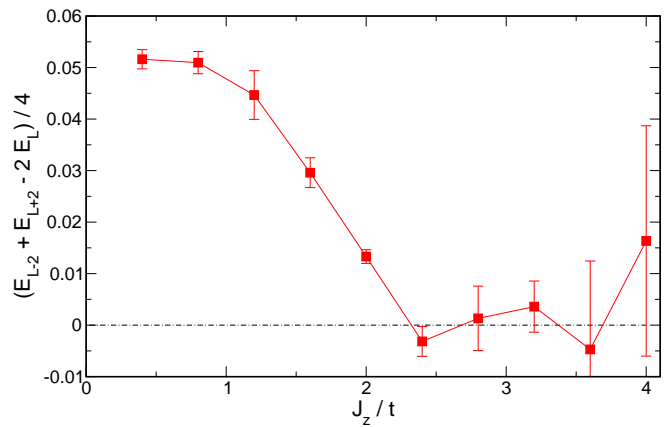}
  \caption{ The estimated two-hole gap in the bulk limit 
    $[E(n_h+L+2,S^z=0)+E(n_h=L-2,S^z=0)-2E(n_h=L,S^z=0)]/2$ at
    quarter-filling, as a function of $J_z/t$.} 
  \label{fig18}
\end{figure}

Figure \ref{fig17} shows the two-hole energy gap
$[E(n_h=L+2,S^z=0)+E(n_h=L-2,S^z=0)-2E(n_h=L,S^z=0)]$ as a function of
$1/N$ at selected values of $J_z/t$. For $J_z/t \leq 2.3$, the gap
appears to scale to a finite value in the bulk limit $N \rightarrow
\infty$, whereas the limit is consistent with zero from there on. Figure
\ref{fig18} shows the estimated bulk limit as a function of $J_z/t$. 
The finite charge gap for $J_z/t < 2.3$ corresponds to a cusp in the energy
density versus filling factor surface, and is
evidence of the formation of a
CDW state at this commensurate filling factor, according to
White {\it et al.} \cite{white2002}.

\begin{figure}
\includegraphics[width=\linewidth]{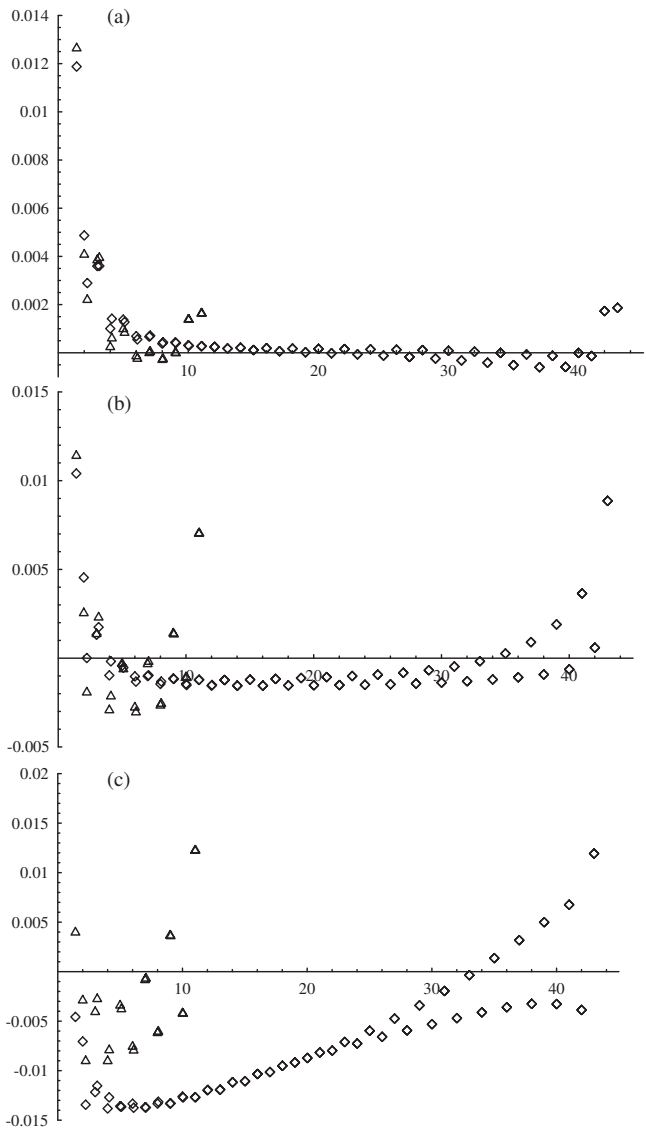}
\caption{ The density-density correlation function as a function of
distance, with one site fixed at the middle of the ladder and the other
varying, at quarter filling. Triangles: $L=22$; squares: $L=88$.
Case a): $J_z/t=2.0$;
b): $J_z/t = 2.6$; c): $J_z/t = 3.0$ }
 \label{fig19}
\end{figure}

We have looked for further evidence of the CDW state in the electron
density profiles and correlation functions. The density profile on an
odd-rung lattice at quarter filling shows some evidence of rung-to-rung
oscillations, but they die away as $N \rightarrow \infty$, and so do not
seem to correspond to the CDW.

Examples of the density-density correlation functions are shown in
Figure \ref{fig19}, with one site fixed at the middle of the ladder, and
the distance to the other site varying, including both sites on the same
leg and those on the opposite leg of the ladder. There is a positive
correlation for nearby sites, and a Friedel rung-to-rung oscillation
near the boundary. In the intermediate region, there is no strong
evidence of either a rung-to-rung
alternation or a checkerboard alternation. 
One major qualitative difference is obvious, however,  between Figures
\ref{fig19}a) and \ref{fig19}c). At $J_z/t = 2$, the correlation
function is virtually zero at intermediate and large distances, as one
would expect of a solid phase:
this is consistent with the CDW behaviour. 
For
$J_z/t = 3$, on the other hand,  the correlation function has a substantial 
dip at
intermediate distances, which actually increases with increasing lattice
size. This behaviour is more reminiscent of a liquid or a gas.

\section{Discussion}
\label{secV}

\begin{figure}
\includegraphics[width=\linewidth]{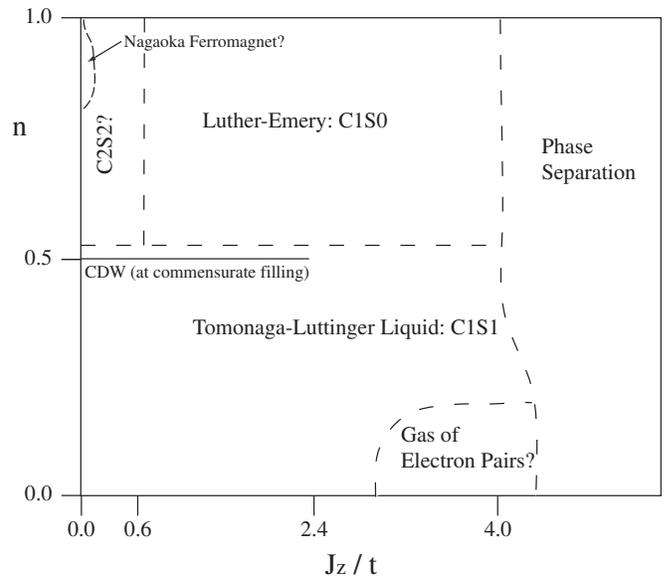}
\caption{ Schematic phase diagram for the t-$J_z$ ladder.}
 \label{fig21}
\end{figure}

As expected, the phase behaviour of the t-$J_z$ ladder at $T=0$ appears
to be broadly similar to its counterpart, the t-J ladder. A schematic
diagram is shown in Figure \ref{fig21}. At half-filling, the system is
equivalent to an Ising antiferromagnet, with long range magnetic order;
but at any finite temperature or hole density, the order parameter will
vanish, leaving only finite-range antiferromagnetic correlations.

Near half-filling, phase separation into hole-rich and hole-poor regions
occurs beyond $J_z/t \simeq 4.0$, which is about half the critical value
for the t-$J_z$ chain \cite{batista2000}, but twice the critical value
for the full t-J ladder \cite{troyer1996}. 
Below that, there is a C1S0 or Luther-Emery phase, where the holes bind
into pairs and the charge gap vanishes, but the spin gap remains finite.
The spin gap is discontinuous at half-filling, because the exciton spin
gap is less than the magnon gap. These features are identical with those
found for the t-J model \cite{troyer1996}. We have not explored the
question whether there are superconducting correlations in this region,
but they are expected to be present if $K_{\rho} > 1/2$.

The binding between hole pairs vanishes in the $S^z=1$ channel below
$J_z/t \simeq 1.8$, and in the $S^z=0$ channel below $J_z/t \simeq 0.6$.
Below that, the holes repel each other, and we presumably enter the C2S2 gapless
phase discussed by M{\" u}ller and Rice \cite{mueller1998} for the t-J
case (although exact diagonalization data up tp $N=14$ are somewhat
equivocal as to whether the $S_z = 1$ states are degenerate).  
These different regimes are clearly evident in the hole density profiles
on a finite lattice.

In the Nagaoka limit as $J_z/t \rightarrow 0$, a finite-lattice
calculation reveals rather different behaviour in the two models in the
one-hole sector. In the
t-J case, a state with maximal spin $S$ crosses over the $S=1/2$ state
to become the ground state as $J/t \rightarrow 0$. For the t-$J_z$
system, on the other hand, the $S^z=1/2$ state remains the ground state
at all $J_z/t$, and the Nagaoka limit is approached smoothly. States
with higher $S^z$ become degenerate with the $S^z=1/2$ state in the
limit, to form a multiplet with maximum total spin $S$. Any
ferromagnetic correlations in the ground state appear to be transverse
to the $z$ direction, as argued by White and Affleck \cite{white2001}.
Our DMRG results for the one-hole gap appear to decrease as
$(J_z/t)^{1/2}$, rather than $(J_z/t)^{2/3}$ as predicted by theory; the
theory needs re-examination for this case.

Some crude estimates of the effective Hamiltonian parameters in the
Luther-Emery phase were made at half-filling. Friedel-type fits to the
density profiles give $K_{\rho} \simeq 0.8-1.0$, although it is not
clear that the Schulz argument \cite{schulz1999} that $K_{\rho} = 1$ at
half-filling is applicable to the t-$J_z$ case. We found that
$v_{\rho}=0$ in that limit, a general result which should apply to
chains and ladders for both models.

We have also made some studies of the quarter-filled case. For $J_z/t
\leq 2.3$, the system appears to be in a CDW state, with a finite
two-hole gap and solid-like density-density correlation functions,
although we could not discover any particular pattern of charge density
waves. 
This behaviour was predicted for a
commensurate filling factor by White, Affleck and Scalapino \cite{white2002}.
We find that the spin gap vanishes in this phase, and in fact is
compatible with zero at all couplings. For $J_z/t \geq 2.3$, any pairing
effects are weak or nonexistent, and the system appears to be in a C1S1
Tomonaga-Luttinger phase
with a single gapless spin and single
gapless charge mode,
similar to that discussed for the t-J ladder at densities at or below
quarter-filling 
by Hayward and Poilblanc \cite{hayward1996}.
The calculations become very unstable beyond $J_z/t \simeq 3$, and we
take this as indicating the onset of phase separation again at larger
$J_z/t$.
At very small electron densities, an
electron-paired phase should exist. These findings are indicated schematically in
Figure \ref{fig21}.

\acknowledgments
This work forms part of a research project supported by a grant
from the Australian Research Council.
We are grateful for extensive computational support  
provided by the Australian Partnership for Advanced Computing (APAC)
National Facility and by the Australian Centre for Advanced Computing
and Communications (AC3).

\end{document}